\pdfoutput=1
\documentclass[ ,final] {aipproc}
\layoutstyle{8x11double}

\begin{document}

\title{The Enigma of B-type Pulsators in the SMC}


\author{S\'{e}bastien Salmon}{
  address={Institut d'Astrophysique et de G\'eophysique de l'Universit\'e de Li\`ege,
All\'ee du 6 Ao\^ut, 17 B-4000 Li\`ege, Belgium}}

\author{Josefina Montalb{\'a}n}{
  address={Institut d'Astrophysique et de G\'eophysique de l'Universit\'e de Li\`ege,
All\'ee du 6 Ao\^ut, 17 B-4000 Li\`ege, Belgium}}

\author{Andrea Miglio}{
  address={Institut d'Astrophysique et de G\'eophysique de l'Universit\'e de Li\`ege,
All\'ee du 6 Ao\^ut, 17 B-4000 Li\`ege, Belgium},
altaddress={Postdoctoral Researcher, Fonds de la Recherche Scientifique - FNRS, Belgium}}

\author{Thierry Morel}{
  address={Institut d'Astrophysique et de G\'eophysique de l'Universit\'e de Li\`ege,
All\'ee du 6 Ao\^ut, 17 B-4000 Li\`ege, Belgium}}

\author{Marc-Antoine Dupret}{
  address={Institut d'Astrophysique et de G\'eophysique de l'Universit\'e de Li\`ege,
All\'ee du 6 Ao\^ut, 17 B-4000 Li\`ege, Belgium}}

\author{Arlette Noels}{
  address={Institut d'Astrophysique et de G\'eophysique de l'Universit\'e de Li\`ege,
All\'ee du 6 Ao\^ut, 17 B-4000 Li\`ege, Belgium}}

\begin{abstract}
Since the early  nineties it is accepted that the excitation mechanism of B-type pulsators on the main sequence is due to the opacity peak in the iron-group elements at $T\approx 200,000$~K. The Fe content plays then a major role in the excitation of $\beta$ Cep and SPB pulsations. While  theoretical non-adiabatic computations predict no $\beta$ Cep pulsators and only a small number of SPBs for low metallicity environments such as that of the Magellanic Clouds (MCs), recent variability surveys of B stars in the SMC have reported the detection of a significant number of SPB and $\beta$ Cep candidates.
Since the SMC is the metal poorest (Z$\approx$0.001-0.004) of the MCs, it constitutes an interesting object for investigating the disagreement between theory and observations.
We approach the problem by calling into question some of the hypotheses made in previous studies: given the different chemical evolution of the SMC compared with our local galactic environment, is it appropriate to describe the chemical composition of SMC B stars by scaling the solar mixture to lower $Z$? Is that composition uniform in space and time?
In this paper we present the results of a stability analysis of B-type stellar models computed with a revised chemical composition and metallicity specific to the SMC.
 \end{abstract}

\maketitle



\section{B stars in the SMC}

\begin{figure}
  \includegraphics[height=.29\textheight]{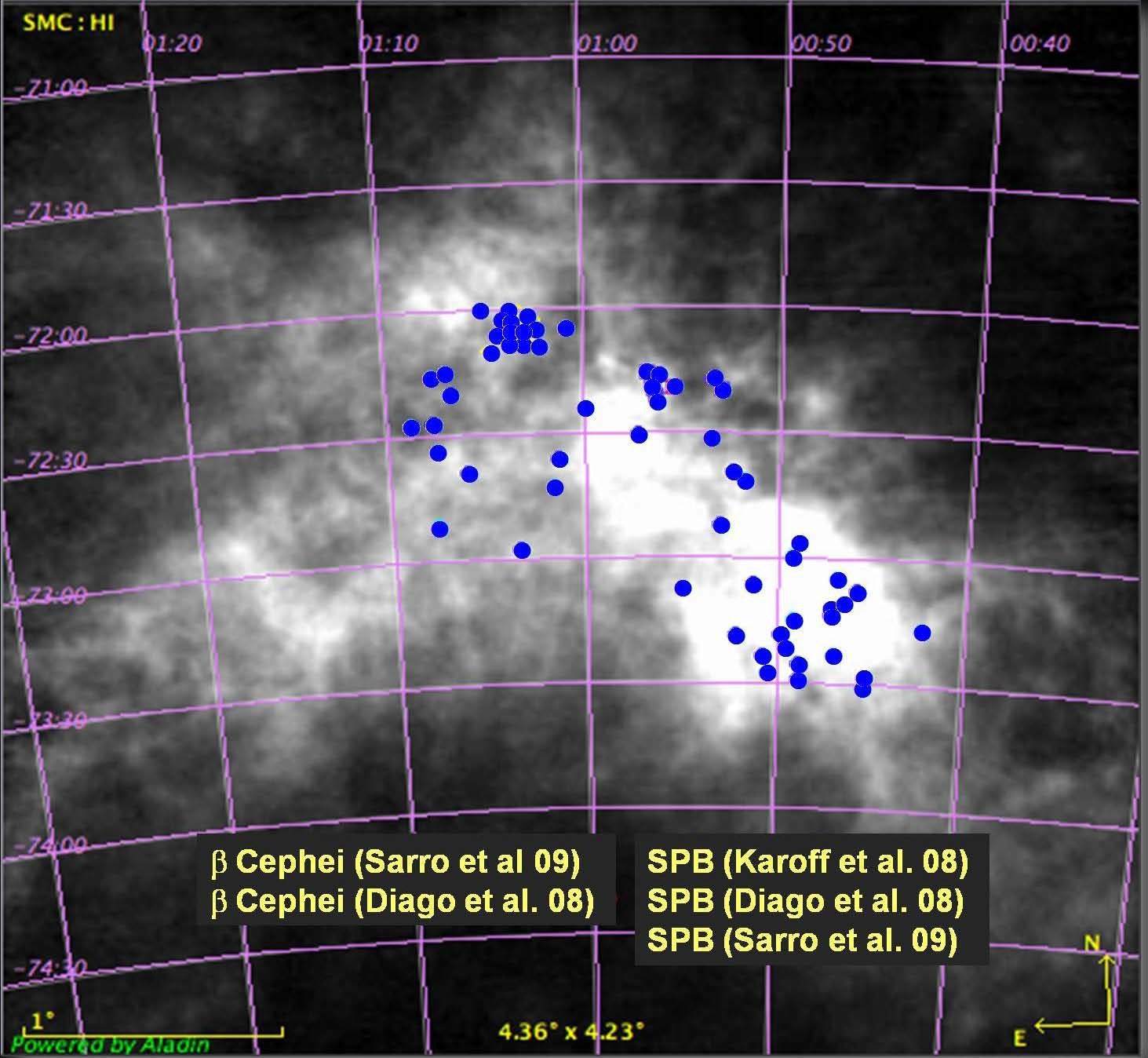}
  \caption{Map of column density of the neutral hydrogen in the SMC. The brighter is the colour, the higher is the density . Colored dots correspond to SPB and $\beta$ Cep candidates. }
\label{mapa}
\end{figure}

During the last three  years the number of B-type candidate pulsators in the SMC has steadily increased. \citep{Ko06} looked for variability in SMC stars by using OGLE-II \citep{So02} and MACHO \citep{MACHO} data, and found six $\beta$~Cep and eleven SPB candidates. Later on, \citep{Sarro} detected 43 multiperiodic B-type pulsators (23 $\beta$~Cep and 20 SPBs)  by automatic inspection of OGLE light-curves. Moreover, observations of two open clusters in the SMC (NGC~330 and NGC~371, \citep{Diago} and \citep{Karoff} respectively)  have also revealed the presence of B-type pulsators (37 SPBs and 1 $\beta$~Cep). In Figure~\ref{mapa} we place these hot pulsator candidates on a neutral hydrogen map of the SMC \citep{Stanimirovic}.  We note that all candidates, either $\beta$ Cep or SPBs,  are located  in the  main bar of the SMC, i.e. the region with the  highest H I density and so with the most active star formation.
This is in good agreement with the fact that main sequence B stars are young objects, with a typical age of $1.3\times 10^{8}$ yrs  and $1.5\times 10^{7}$~yrs for 4~M$_{\odot}$ and 12~M$_{\odot}$ respectively.

\subsection{Chemical abundances in the SMC}

Recent studies have determined an age-metallicity relation for the SMC  based on star clusters \citep{Carrera} and planetary nebulae \citep{Idiart}, and both concluded that  a clear increase in the metallicity occurred during the last 2 Gyr. In particular, we expect [Fe/H]$\sim -0.7$ for young objects such as B stars, wherever they lie in the galaxy \citep{Carrera}.

\begin{figure}
  \includegraphics[width=.45\textheight]{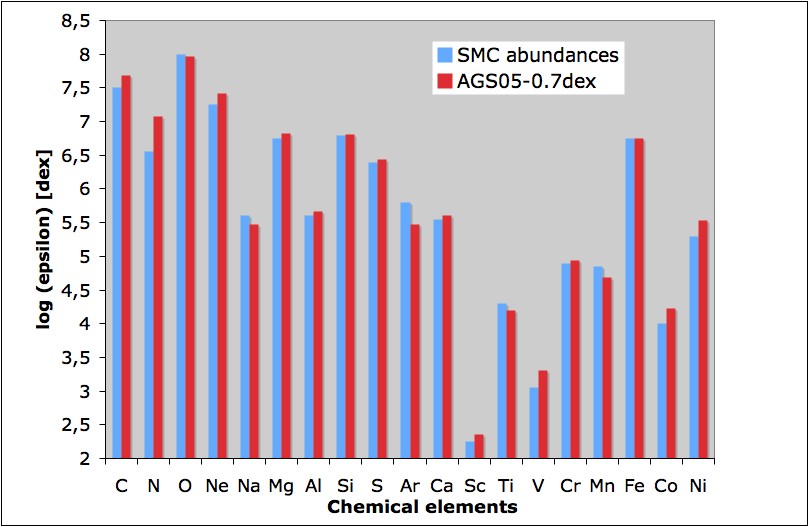}
  \caption{Element abundances (mean values) in B-type stars of the SMC, compared to abundances from AGS05, scaled down by 0.7 dex, i.e. to Z\ =\ 0.0026. Log $\epsilon$ gives the abundance in the scale $\log(N_x/N_H)+12$. Note that the $y$-axis starts at 2.}
\label{abun}
\end{figure}

Since the efficiency of the $\kappa$-mechanism to drive pulsations in $\beta$~Cep and SPBs strongly depends on the abundance of the iron-group elements, the detailed metal mixture of the SMC may be relevant for a stability analysis of hot stars in this galaxy. We have assembled a set of representative abundance values for young massive stars in the SMC from a comprehensive and critical examination of literature data for H II regions, B stars and cool supergiants in the same mass range. The reviewed elements are C, N, O, Ne, Na, Mg, Al, Si, S, Ar, Ca, Sc, Ti, V, Cr, Mn, Fe, Co and Ni from which we obtain a metal mass fraction $Z=0.0024$.  The mean value of the iron abundance, $\log \epsilon(Fe)=6.75$, is 0.7~dex lower than that in the solar mixture  \citep[thereafter AGS05]{Asplund}. There is a high dispersion of the Fe abundance ([Fe/H]$\sim -0.7\pm0.3\ $dex) but that probably comes from different data treatments used in the works we rely on.

\begin{figure}
  \includegraphics[height=.35\textheight]{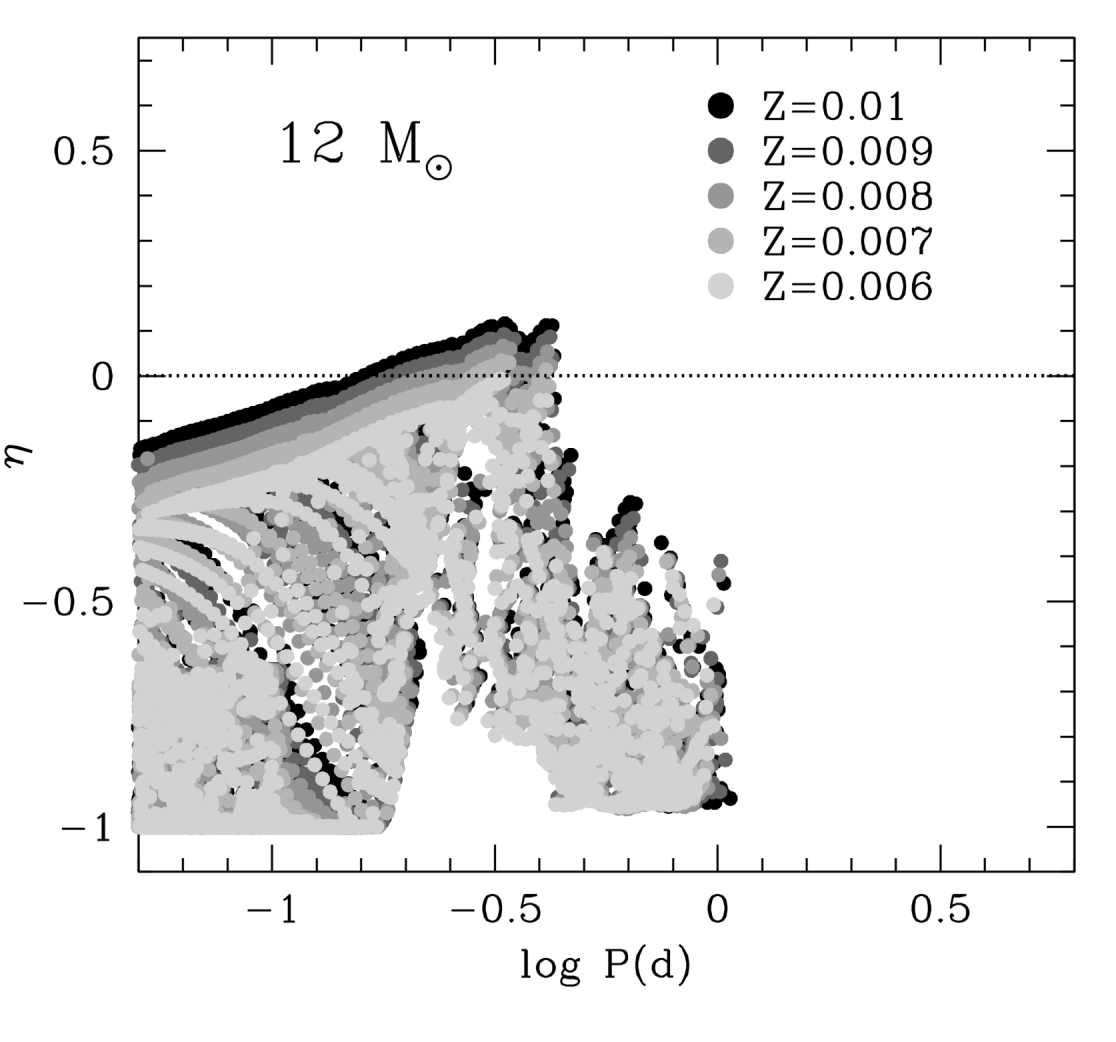}\includegraphics[height=.35\textheight]{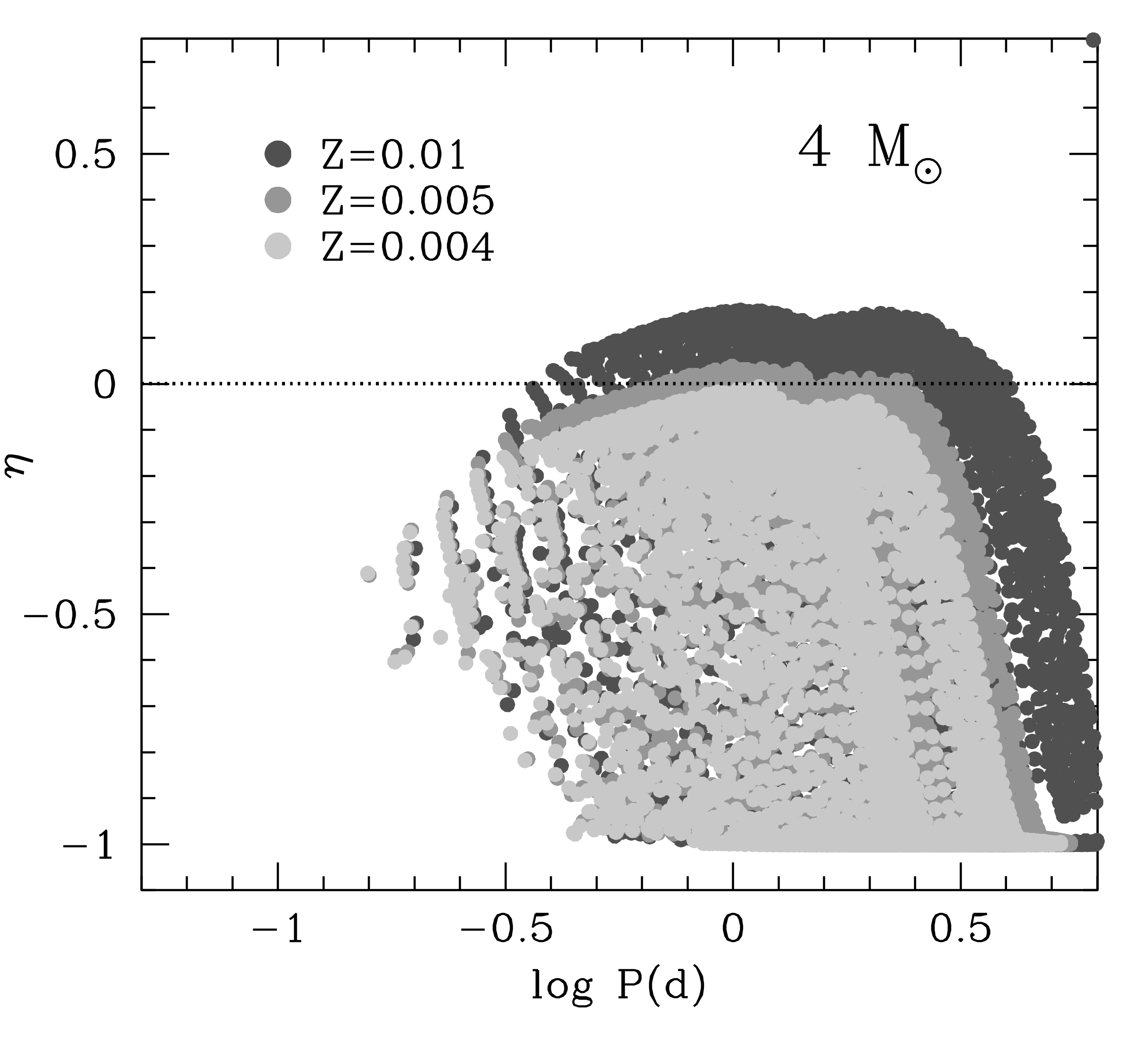}
  \caption{Measurement of instability of  $\beta$ Cep (left panel) $\ell=0$, 1 and 2 modes,   and SPB (right) high order $\ell=1$ and 2 g-modes as a function of the mode period ($\log P$ (d)). $\eta > 0$: unstable mode, $\eta < $ 0 stable mode. Models along the main-sequence  evolution of 12 and 4~M$_{\odot}$  for different values of the metal mass fraction $Z$ and the SMC B stars metal mixture are considered.}
\label{bandas}
\end{figure}

The iron abundance we determine is thus in good agreement with the value predicted for young objects by the age-metallicity relation in the SMC.
In Figure~\ref{abun} we compare the new metal mixture of SMC B stars with the solar abundance of AGS05 scaled by 0.7~dex. We note that there is no obvious difference, except for N, between these two mixtures. We do not expect a large enough increase in the iron-group element opacity bump to excite pulsation modes in B stars.

\section{B-type stars instability in low metallicity enviroment}

We computed new opacity tables for the "SMC B stars mixture" by making use of the OP opacities, \citep{Badnell}. They are more suitable for our aim as they are the ones taking into account the largest number of ionic/atomic transitions for iron. We used these tables to model stars with masses from 2.5 to 18 $M_{\odot}$, for $Z=0.0024$, 0.005 and 0.01, keeping the helium mass fraction fixed at $Y=0.28$. The overshooting parameter was set to 0.2 pressure scale height, we used the Eddington's law ($T[\tau]$) as surface boundary condition,  and OPAL equation of state \citep{Rogers} was employed. The stellar models were computed  with the stellar evolution code CL\'ES \citep{cles}, and the non-adiabatic pulsation analysis was performed with MAD, \citep{Dupret}.

As expected from the similarity between the SMC B stars mixture and the solar one scaled down by 0.7~dex, our results are quite similar to those obtained by \citep{Miglio3}: we do not obtain excited pulsation modes for $Z=0.0024$, neither in $\beta$~Cep nor in SPBs. For $Z=0.005$ we get only SPB excited modes. For $Z=0.01$ both the instability strips of  $\beta$~Cep and SPB in the HR diagram and the frequency domain of excited modes  are very close to those obtained by \citep{Miglio3} using AGS05 solar mixture.

 To derive the minimum $Z$ required to excite B-type pulsations, we have also performed  non-adiabatic analysis of main-sequence stellar models with 4 and 12~M$_{\odot}$ (in the middle of the SPB and $\beta$~Cep instability strips respectively)  and  metal mass fraction ranging from 0.003 to 0.01 with a step of 0.001. The results are shown in Figure~\ref{bandas} where we plot the quantity $\eta$, a normalized growth rate \citep{Stellingwerf79}. $\eta > 0$ corresponds to an unstable mode and $\eta < 0$ to a stable mode, and  tells us how far from being excited the pulsation mode is. We note that $Z \geq 0.007$ is required to excite $\beta$~Cep modes with the SMC metal mixture, and $Z > 0.004$ for SPB mode excitation.

\section{Conclusions}

Stellar models with the chemical composition representative of B stars in the SMC do not show excitation of SPB nor $\beta$ Cep modes. For values of $Z \leq 0.004$, no mode is excited.  The apparent disagreement between observations and theory  remains, and raises the following questions:\\

\begin{enumerate}
\item Does the mean chemical composition of SMC B stars represent the one of  B-type pulsators?
\item Is the dispersion in the Fe abundance  only an effect of measurement errors? Could it be at least partially real?
\item Is the opacity in the iron-group element bump underestimated?
\end{enumerate}

These questions open new perspectives for investigating this problem in the future.


\begin{theacknowledgments}
JM acknowledges financial support from the Prodex-ESA Contract Prodex 8 COROT (C90199).
This work has made use of OP opacity server (\textit{http://opacities.osc.edu/}) and Aladin (\textit{http://aladin.u-strasbg.fr/}) web tools.

\end{theacknowledgments}



\bibliographystyle{aipproc}   

\bibliography{santafe_salmon.bbl}

\IfFileExists{\jobname.bbl}{}
 {\typeout{}
  \typeout{******************************************}
  \typeout{** Please run "bibtex \jobname" to optain}
  \typeout{** the bibliography and then re-run LaTeX}
  \typeout{** twice to fix the references!}
  \typeout{******************************************}
  \typeout{}
 }

\end{document}